# STRATEGIES FOR THE DETECTION OF ET PROBES
## Within Our Own Solar System

**JOHN GERTZ** Zorro Productions, 2249 Fifth St., Berkeley, CA, 94710, USA

**Email**: jgertz@zorro.com

Arguments are reviewed in support of the hypothesis that ET would more likely send physical probes to surveil our Solar System and communicate with Earth than to communicate from afar with interstellar radio, infrared or laser beacons. Although the standard SETI practice of targeting individual stars or galaxies with powerful telescopes might detect a foreground local probe by serendipity, an intentional hunt for those probes would entail a different set of strategies, most notably sacrificing sensitivity (needed to detect a very faint and very distant signal) in exchange for a widened field-of-view (because a local signal can be reasonably hypothesized to be relatively bright). This paper suggests a number of strategies to detect local ET probes.

**Keywords:** SETI, ET, Probes, Lurkers, Nodes, METI

## 1 INTRODUCTION

The first paper describing a modern method for the detection of extra-terrestrials (ET) was published by Cocconi and Morrison in 1959 [1]. Morrison and Cocconi had proven that a strong radio transmitter could send decodable signals across interstellar space (beacons). The next year, Frank Drake conducted the first modern SETI search when he trained the Green Bank Telescope (GBT) on two nearby Sun-like stars [2]. In 1961 Schwartz and Towne expanded the scope of SETI with a proof that powerful lasers could also transmit information across interstellar space [3], thus giving birth to optical SETI (OSETI). Less well-remembered, but also in 1960, Robert Bracewell first suggested that aliens might prefer to send surveilling and/or information laden probes to our Solar System rather than to communicate via electromagnetic (EM) interstellar beacons [4, 5]. Since then, others, as well as myself, have enhanced the arguments favoring local AI probes over interstellar beacons [6-13].

Among those reasons are the following:

- Probes might outlive their progenitor civilizations, diminishing or eliminating the need for an L factor in the Drake Equation, which posits a dependence between the likelihood of a detection and the average lifetime of alien civilizations. It stands to reason that if a civilization only transmits signals from its home star for 10,000 years, and that those 10,000 years do not overlap with our current time (as adjusted by distance in light years), then we will not be within that signal's light cone and therefore not be able to detect its signal. However, probes might outlive their progenitor civilizations by an arbitrarily long period of time. Even if two civilizations co-exist in time, their transmitters and receivers also must align in time. Unless one or both have all-sky-all-the-time capabilities or at least a very short duty cycle, temporal alignment will not likely be achieved.

- Probes can surveil their target solar systems, and therefore can conduct useful science on behalf of the civilization(s) that sends them. Beacons return no information unless and until detected by a technological species and a message deliberately sent in reply. Flyby alien probes might have determined that Earth was biologically active billions of years ago. Perhaps they have continued to fly by at 100 million-year intervals ever since, noting the rise of atmospheric oxygen and the biosphere's advance to multicellularity. Perhaps, as an example, in accordance with ET's models indicating the average length of time from the onset of multicellularity to the emergence of the first technologically capable species, permanent probes were first placed into our solar system 100 million years ago.

- A distant radio beacon's signal may be so diminished by the time it reaches Earth that in the absence of a gargantuan radio receiver and long integration times, nothing more than a carrier signal may be discernible. A laser would fade to nothing due to signal spread and absorption by interstellar dust within approximately 1,000 light years. Because the transmitting civilization may be vastly different than the receiving civilization, the former's message, even if accurately received, may not lend itself to decoding, such that we will never understand what was intended. The situation with a local probe is completely different. Once a technological civilization does arise on a surveilled planet, after a period of study and decoding, the probe is perfectly positioned to communicate to that civilization with a flux that will be orders of magnitude greater than a distal beacon, and in a manner that will be comprehensible.

- Probes are safer than beacons for their progenitors. In the absence of any knowledge of a target civilization's capabilities and intentions, probes offer anonymity, as they need not reveal their progenitor's location. Beacons, virtually by definition, do.

- Probes throughout the Milky Way might be linked by way of a communications network of nodes with knowledge constantly accumulating and growing within the network, regardless of whether the total number of currently contributing civilizations increases or decreases with time. The fact that Earth is a biological planet may be recorded in an atlas appended to the so-called Encyclopedia Galactica, and therefore known to all, regardless of whether any particular civilization has analyzed Earth's atmosphere for itself because it just happens to have an edge-on line of sight to Earth as it transits the Sun.

Although speculation about ET probes is as old as speculation about interstellar ET beacons, almost all search efforts to date have been focused on the detection of interstellar beacons, and almost none have explicitly hunted for local probes. Given the strong arguments favoring probes, a possible reason for this imbalance can be framed by the old expression, "if all you have is a hammer then every problem becomes a nail." Applied to SETI, astronomers armed with such sensitive radio telescopes as Arecibo and Green Bank, and such powerful optical telescopes as Keck, failed to search for local and potentially far more obvious targets for which such sensitive telescopes would not be necessary. Although such astronomers can sensibly argue that their interstellar searches might detect a foreground probe by serendipity, it is equally true that a search for local probes might detect by serendipity a background interstellar beacon. Nonetheless, best search strategies between the two paradigms will differ.

## 2 ARE PROBES LURKERS?

Allen Tough was the first to adopt the term, "lurkers," in the nonfiction SETI literature to describe probes that were intentionally hiding from us within our Solar System. In the 1990's he ran a welcoming website where ET might announce itself. David Brin popularized it, proposing that they are intentionally hiding within our Solar System. In an open letter, David Brin made a passionate plea to these lurkers to reveal themselves [14]. In doing so, he gave many reasons why lurkers might be lurking; for example, they are stealing our IP without any intention of giving us their knowledge in return, they are waiting until we have achieved some further developmental milestone, or they wish to spare us from culture shock. Brin's denotation of "lurkers" has been taken up by others [15]. Clearly, the term "lurkers" is a pejorative, conjuring images of pedophiles or muggers, or, more generally denoting nefarious purposes.

While Brin's various scenarios are plausible, he neglected to include the four most likely reasons as to why we have not yet detected an ET probe:

- Prior to about 100 years ago, an ET probe's camera might have registered some indication that one Earth species had developed some technological capabilities, perhaps by detecting agriculture, pyramids, or cities. However, only the commencement of EM transmissions would have afforded an ET probe the possibility of in-depth study of the nature of that species. ET may now be surveilling Earth in an attempt to understand us before initiating communication. Thanks to the arduous decades-long research of E.O. Wilson and others, we now know something about the lives of ants [16]. However, we still know next to nothing about the communications of cuttlefish, dolphins, or ravens, and are far even from a clear understanding of chimpanzees. It may be that understanding human beings and their languages is just too difficult for an ET probe to decode in only the roughly 100 years since it might first have detected artificial electromagnetic waves (EM) emanating from Earth, even allowing that such a probe might have an onboard computer of immense power. Moreover, from the probe's point of view, until the early 50s and the beginning of regular television broadcasts, it would have had to rely only upon a cacophony of disembodied radio voices in a multiplicity of languages, lacking all visual referents. It is hard to imagine what useful information might be gleaned from that. Perhaps an alien probe is still trying to decipher our language through Sesame Street, our math through Khan Academy, and the rest through YouTube. We will hear from such a probe only after it has decoded our languages and our culture. However, when we do hear from it, the transmission will be bright, and its transmission may be in the Queen's own English, or, alternatively, in Chinese, Urdu, or some other terrestrial language. A sub-possibility is that the probe is hibernating, perhaps buried beneath the regolith of an asteroid for protection from radiation and micrometeorites, and only pokes out an antenna once a century or once a millennium to test for artificial EM. Presumably, it knows that Earth is biologically active, but would not know exactly when Earth's first technological species might emerge. In such a case, it may not have been monitoring Earth's EM since the first possible moment about a century ago, that is, with the advent of commercial radio broadcasts.

- The probe may not be fully autonomous but reports back to some home base. I have postulated that the galaxy may be laced with a communication network mediated by and through nodes that might be located around many or most stars [10,11,12]. For the purposes of this paper, probes will be considered as those alien crafts that are surveilling and possibly now or in the future transmitting to Earth, while nodes will be considered to be further away (e.g., at 550 AU) communication relay hubs. The local probe(s) would transmit its Earth-related data back through the communications network to its home base for fuller decoding and analysis, as well as for instructions on how to proceed. If such a local probe is in communication with a home base, say, at a distance of 100 light years, and that data stream began in about 1920, then the very earliest that we might expect to hear from that probe would be 100 years from the present.

- As a consequence of the foregoing two points, ET may be on the cusp of communicating to Earth with a loud or bright signal emanating from within our own Solar System. The timeframe for the arrival of that first communication is unknowable, but may be measured in months or years, rather than centuries or millennia. It is incumbent upon Earth to be as ready as possible to receive that signal.

- The local ET probe(s) is communicating with us, but we have failed to detect the communication simply because we have not pointed telescopes at the right coordinates, or we have detected it, but have mischaracterized and therefore ignored the signal. If a radio signal, it might have been miscategorized as RFI, for example, because of the lack of an expected Doppler shift (i.e., the local ET probe might correct for its motion and the Earth's orbital and rotational motion, but probably not for latitude differences in rotational motion due to the different latitudes of the various telescopes that might first detect its signal), or because, being in local motion, it would be absent upon re-targeting the same sidereal coordinates. If it was a laser, it might not have been subjected to spectrographic examination and found to be monochromatic, and hence artificial.

- A fifth possibility, not mentioned by Brin but nonetheless plausible, is that galactic metalaw [17] stipulates that it is incumbent upon a primitive civilization, such as our own, to initiate contact if we want it. This has been the contention of a number of SETI researchers and enthusiasts for some time, some of whom have organized as METI International [18]. This is known as Active SETI or Messaging SETI (METI), and it is highly controversial [19, for a review].

Since none of these reasons are nefarious in and of themselves, the term "lurkers" should be avoided, and researchers are cautioned to revert to Bracewell's neutral term, "probes."

## 3 SEARCH STRATEGIES

Each of the five contingencies cited above suggests a set of search strategies. While assigning search techniques to each contingency, it should be emphasized that there is overlap among them, such that a technique suggested by one contingency might be effective for another as well.

### 3.1 ET is surveilling us with a fully autonomous probe(s) that may communicate in due course when it is ready or may remain silent indefinitely (i.e., it is truly lurking).

An ET probe operating under this eventuality will be the most difficult to detect. Not only will it not be transmitting to Earth at this time, but neither will it be in communication with any other ET probe or node, making it subject to possible eavesdropping. However, once that fully autonomous local probe does intentionally signal Earth, reason suggests that it will be an unmistakably bright or loud signal relative to the capabilities of our most sensitive telescopes.

Until then, our two best hopes for detecting a non-transmitting probe are (a) to look for an anomalous optical object or (b) to detect its infrared (IR) waste heat.

1. The most obvious vantage point from which an ET probe might surveil Earth would be from Earth orbit. A thorough examination of existing public and classified databases is warranted. Perhaps an alien probe has been misidentified as space junk. Perhaps there is a satellite that the U.S. assumes is a secret Russian or Chinese spy satellite and that the Chinese and Russians assume is an American classified spy satellite, but is in fact alien. Perhaps an ET probe is hiding behind a human built satellite [20]. However, other than for purposes of camouflage, there is no apparent reason why an alien probe need be in either a low Earth orbit (why would it submit itself to atmospheric drag?) or in a geosynchronous orbit, so a search for an Earth orbiting probe should be expanded to encompass all directions and altitudes. Whatever its orbit, it would likely require some sort of propulsion system to prevent long term orbital degradation.

2. Perhaps a probe was in Earth orbit until the mid-20th century and then decamped to an asteroid or some other place to avoid detection after we became spacefaring. It may be that an object assumed to be an asteroid is in fact an alien probe. The entire database of asteroids should be examined under this hypothesis for anomalies such as an unexpected IR signature, albedo or trajectory that does not strictly obey the laws of gravity [21]. Unfortunately, this search strategy would currently apply only to either a large probe or a small probe that is in close proximity to Earth. The most sensitive asteroid detector, Atlas, can only detect a 10 m wide ET probe within ~ 0.04 AU, with sensitivity dropping dramatically when looking at twilight, i.e., at the Solar System within 1 AU [22]. However, the Vera Rubin Observatory (formerly, LSST), will soon commence its 10-year mission which will greatly increase the number of known main belt asteroids and near-Earth objects. Barring stealth technology, a detectably large ET probe will have a measurable albedo. The problem is how to distinguish this albedo from that of an asteroid. Asteroid albedos range from 0.03-0.5. Object size is deduced from the albedo. That part of the solar flux which is not reflected as albedo is absorbed and reradiated as IR. A probe, unless fully hibernating, will radiate an added IR waste heat signature due to its internal computations and other operations. However, since estimates of object size are always subject to wide error bars, and they are correlated with albedo and IR, the added IR effect would have to be quite pronounced to be judged anomalous (on the order of hundreds of MW). There may be other indications that a putative asteroid may in fact be artificial, such as an anomalous shape or anomalous surface composition; or perhaps a group of anomalous objects share a discernable pattern in their relative motions [23].

3. It is possible that an alien probe is surveilling Earth, not from Earth orbit but from a lunar orbit or from the surface of the Moon [24]. There are advantages to being deployed on the surface of the Moon, as it would allow the probe access to raw materials for repair or the build out of capabilities, and to burrow into the regolith for protection from radiation and micrometeorites. On the other hand, relative to an orbiting probe, a landing probe would require additional hardware devoted to that purpose. The Lunar Reconnaissance Orbiter's (LRO) Narrow Angle Camera has been imaging large swaths of the Moon since 2009 at 0.5 m/pixel resolution. Davies and Wagner have suggested that this vast photographic library should be analyzed for traces of an alien probe, or any other alien object such as trash or structures constructed for purposes other than surveillance. The best way to do this might be through a large-scale citizen scientist program on the model of Galaxy Zoo where volunteers would examine successive photos for any sort of anomaly, including unusual colors, reflections or shapes. Although automated searches might be possible, since we do not know exactly what to look for, it is hard to write software to cover all contingencies, though machine learning techniques might be used to overcome this objection [25]. The moon may be less desirable as a landing spot relative to an asteroid, as its surface may have lower quality raw materials to work with. However, its great advantage relative to asteroids is its much closer proximity to the object of its surveillance, Earth. From the point of view of SETI scientists, it would be vastly easier to examine the surface of the Moon for alien probes than almost any asteroid, the exception being those very few that have been visited and photographed by such spacecraft as Rosetta, Dawn, and Osiris-REx. Closeup photos of asteroids visited by space missions should be subjected to the same examination for anomalies as those of the Moon.

### 3.2. ET's probe is not signaling Earth, but is in communication with a node, perhaps around nearby stars

1. A nodal interstellar communication system may achieve signal gain through the use of gravity lenses. One problem facing my own intergalactic nodal hypothesis has been that the local probe must possess a very powerful radio or laser transmitter, or the receiving node must possess a very large antenna or telescope, or both. However, there is a distance around every star from which a transmitter might achieve enormous gain by utilizing the gravity of the star to lens the signal in accord with Einstein's Theory of General Relativity. The focal line of the solar gravitational lens (SGL) for our Sun begins at 550 AU and extends outward from there [26]. Michael Hippke has suggested that an inner Solar System probe surveilling Earth might relay information to a node that resides on the SGL line and at a point that is in direct opposition to a node that is perhaps located in a nearby star system. EM transmitters would achieve gains of order $10^9$ over direct probe to probe transmissions, and increase data rates by order of $10^6$ [27, 28]. These relay nodes cannot be in Keplerian orbits, or they would very quickly leave prime focus, They therefore must have some propulsion system to allow them to either hover or travel along the focal line. In effect, the intergalactic nodal system would be based upon pairs, with a surveilling probe located within the inner solar system coupled to a communicating node located at the gravity lens focal line. We can calculate where a node(s) is located in our Solar System by calculating those points on the celestial sphere that are >550 AU from the Sun and form a direct line through the Sun to a nearby star. It follows that we might intercept a communication stream when the Earth happens to lie between a local probe and a local node (i.e., local in the sense that it is within our Soalr System, albeit it at >550AU) or between a local node and the non-local node with which it is in communication that is associated with a nearby star system. In theory, this would entail observing those nearby stars where a straight line can be drawn from the SGL's focal line through the star to Sol and the Earth. This would only occur twice a year when the Earth passes through such a SGL line in its orbit. When the intended signal interception is node-to-node, this observing strategy is agnostic as to the location of the surveilling probe, be it on an asteroid, in close solar orbit, in Earth orbit, or elsewhere in the Solar System. Unfortunately, this will not work for the simple reason that there are no nearby stars lying precisely on the ecliptic, the only area in which a star would form that line with the Earth and a Solar SGL. However, there are some nearby star systems that are close enough to the ecliptic that an ET observer located within them will see Earth transit the Sun. 20 such K and G dwarf stars at distances of 7.4 – 143.5 parsecs from Earth have been observed for artificial radio EM in C-band (3.95-8.00 GHz.) by the GBT [29]. Nevertheless, space missions that traverse such points, perhaps piggybacking on an entirely different main mission, may be the best way to test the hypothesis, especially for narrow beam lasers. For example, the ESA Solar Orbiter might just as well have been placed into a solar orbit that traversed an SGL → Sol → nearby star line.

2. In section 3.4.1 below it is argued that a probe may reside in a close Solar orbit. Allowing for this possibility, the area around the Sun should be observed when nearby stars are in opposition to the Sun from Earth's perspective. Observations should also be made of the star that is in opposition to the Sun. This would not seem to be necessary if ET is transmitting a radio beacon from a nearby exoplanet to a probe located in the vicinity of the Sun, because its beam will have in any event widened to encompass the Earth by the time it arrives. But a laser signal from a nearby star would probably be sufficiently narrow such that it would illuminate only a small radius around the Sun if aimed directly at its center. For example, a 10 m laser (i.e., Keck-class) from Alpha Centauri would create a beam width around the Sun of only 0.013 AU (~2 million km).

3. Observations might be made of the SGL lines in our Solar System that point to nearby stars, that is, at a point that is in direct opposition to each nearby star. We would not intercept messages between that SGL and the nearby star, since Earth would not be in that pathway. However, we might intercept a message from the node located at one of these SGL lines to a probe in the inner Solar System. These SGLs should be especially observed when they are close to Solar opposition relative to the Earth as we might be able to eavesdrop on communications from the node in the SGL focal line and a local probe in close solar orbit. The area around the Sun can also be observed at these times to intercept the same train of communication, though of course it is easier to look away from the Sun than toward it [30]. The SGL lines residing in opposition to nearby stars have more importance yet. It is entirely possible that a node positioned in one of these is intentionally signaling directly to Earth rather than directing its communication through an inner Solar System probe. The node may be doing us a favor, as it would reason that an intelligent species will observe the SGL, whereas, as the rest of this paper demonstrates, we can have no such *a priori* certainty as to the coordinates of a local probe.

4. General observations in opposition to the ecliptic and particular observations of nearby stars when they are in opposition to the ecliptic might intercept sister node transmissions to a local probe residing on an asteroid, planet or moon.

### 3.3. ET is not yet signaling Earth, but will in the future with a transmitter that is very powerful relative to what might be expected from the flux of an interstellar beacon

In this event, if the transmission is a radio signal, Earth may already be fully prepared, as there are thousands of mid-size radio dishes all over the world tracking satellites and performing other functions, and less sensitive commercial radio and TV antenna are ubiquitous. If the signal is an optical laser, we can prepare for this eventuality by deploying all-sky-all-the-time capabilities, that is, systems with extremely wide fields of view (FOV), but presumably with correspondingly lower sensitivities. Two such efforts are underway to detect ET lasers pulses, and one is contemplated for wide FOV radio astronomy.

#### 3.3.1. PanoSETI

Shelly Wright, et. al. [31] are in the advance stages of developing and deploying two privately funded Panoramic Optical and Near-Infrared SETI (PANOSETI) instruments at the Palomar and Lick observatories. Two geodesic domes will each house 100 modules comprised of a relatively inexpensive Fresnel lens coupled to an avalanche photodiode (APD) detector. Each lens will have a FOV of 9 square degrees, with 20 arcminute coverage per pixel in the optical and 82.5 arcminutes in the near IR. >8,500 square degrees instantaneous sky coverage will be achieved. Because stars are quite dim at nanosecond levels an alien pulsed laser <30 milliseconds will readily stick out. Although the system is not disadvantaged in the domain of sensitivity, the tradeoff for such a wide FOV is acuity. One dome can localize a putative signal to only 20 arcminutes. However, when triangulated with a simultaneous hit in the second dome at 700 km distance, localization can be reduced to 2-3 arcseconds. Two additional domes are contemplated for the Southern Hemisphere to obtain all-sky capability.

#### 3.3.2. Laser SETI

The SETI Institute has deployed a novel camera system, called Laser SETI, for detecting transient signals over a very wide FOV. The first two systems have been deployed in Northern California, with the intention of deploying additional systems at other sites around the world. Each system is comprised of two 1.5 cm cameras, whose CCDs use Time-Delayed Integration (TDI) to smear out the sky, leaving transient laser signals to stand out [32].

#### 3.3.3. Wide FOV radio SETI

In the event that directed radio beacons are both rare and transient, the most viable way to detect them is with an all-sky-all-the-time telescope. There are quite a few transient radio phenomena for which the widest possible FOV is desirable, including the largely unexplained FRBs. Wide FOV radio astronomy was pioneered by the Survey for Transient Radio Emission (STARE) and formed from crossed dipoles in cavity-type receivers located at three geographically removed sites. These dipoles are temporally aligned such that a simultaneous event recorded in each receiver would be regarded as of astronomical origin [33]. An aperture array telescope has been proposed for SETI and other transient events. This Mid-Frequency Aperture Array Transient and Intensity-Mapping System (MANTIS) would be built adjacent to the SKA-1 in South Africa. It would consist of many thousands of individual antennas, each sensitive to almost the whole sky at L-band, and whose signals would be combined and beam formed in the back end [34].

### 3.4. An ET probe(s) is signaling Earth now but we have failed to make the detection

These techniques are best deployed to detect an ET signal that is less than obvious, but nonetheless of a flux that is higher than what might be expected from an interstellar beacon:

1. One key question is what to target. Although some possible targets are suggested below, it must be admitted that an alien probe might be in any orbit around either the Earth or the Sun. Consequently, the most productive strategy might be to target the entire sky, but with more sensitivity than can be achieved with the all-sky-all-the-time techniques reviewed in section 3.3 above. This would be nearly impossible were the most sensitive and powerful telescopes devoted to the task. For example, if the Keck telescope were to examine each point in the sky for alien lasers for a mere one minute each it would take hundreds of thousands of years to complete a single duty cycle. The Arecibo radio telescope has a total beam size of about 0.002 square degrees at 1.5 GHz, so it would require, in principle, about 20M pointings to mosaic the entire sky (although in practice Arecibo can only image 31.7% of the sky). Allotting 10 minutes to each pointing gives a total duty cycle of about 380 years. However, if an ET probe is intentionally beaming a signal to Earth it can be safely presumed to be powerful enough to be reasonably bright. Breakthrough Listen, using the Green Bank Telescope, observed the recently-discovered interstellar object ‘Oumuamua while it transited our Solar System, down to a sensitivity of 0.08 Watts [35]. ‘Oumuamua is an object of at least the size of an aircraft carrier. If ‘Oumuamua actually were an ET probe or worldship (i.e., populated by alien beings) of such a large size, one can only imagine that it would be able to muster enough onboard energy to transmit at a higher power than 0.08 Watts.

2. It is therefore advised that such a search for a local probe intentionally transmitting to Earth can safely sacrifice sensitivity in favor of widening the FOV. The Allen Telescope Array (ATA) is a far less sensitive radio telescope than the GBT, but it could nonetheless easily detect the 23-Watt signal from Voyager at a distance of 106 AU, using just 10 of its 42 dishes. The ATA could mosaic the approximately 75% of the sky available to it in about 100,500 pointings. If 10 minutes were allotted to each pointing, a duty cycle could be completed in a mere 729 days. This may then be an ideal mission for the ATA.

3. VERITAS is a group of four 12 m. optical telescopes with a 12 m focal length, giving it the very wide FOV of 3 sq. degrees. Each telescope is equipped with 499 PMTs. Originally designed to detect Cherenkov light generated by gamma rays as they hit the atmosphere, VERITAS astronomers took advantage of its ability to detect nanosecond optical events to observe KIC 8462852 (Tabby’s star) for pulsed lasers [36]. Breakthrough Listen has now partnered with VERITAS to broaden its optical SETI capabilities [37]. VERITAS’s large FOV recommends it to all-sky mosaicking.

4. Amateur class telescopes, equipped with either photomultiplying tubes (PMTs) or APDs might mosaic the sky in search of local alien lasers. This would require central coordination, and possibly financial subsidies for participating amateurs from Breakthrough Listen or some other funding source. The telescopes might be grouped into cohorts of ten each, allowing that some will be under cloudy skies, or otherwise not operating on a given night. Each telescope would be controlled by a principle investigator over the Internet, who would be able to slew them from FOV to FOV, perhaps with 10-minute dwell times. Multiple cohorts could observe different FOVs at the same time, such that the entire sky might be observed fairly quickly, depending on the total number of cohorts. Each telescope would be equipped with three PMTs or APDs; each telescope owner would be blind to the identity of the other members of the cohort; and each cohort would be observing the same coordinates at the same time. A single telescope recording simultaneous photon hits within each of its three PMTs or ARDs in any single nano-second could be interpreted as the detection of a laser. Stronger evidence would entail the recording of simultaneous photon hits in multiple nano-second intervals. The strongest evidence would entail simultaneous hits by multiple cohort members. Indeed, a lack of confirmation from other observing telescopes would null the hypothesis of an alien detection.

5. Geoff Marcy is currently searching for local (or background) lasers in the ecliptic, the area within 60 degrees of the Sun (= 0.87 AU), as well as L5 at twilight, by taking successive spectra of large (2 degree square) tiles of the sky at quarter second intervals and analyzing them spectrographically for monochromatic points of light [29].

6. Lacki describes methods for detecting reflections from an ET mirror located within our Solar System that is pointed directly toward Earth [38]. Such a low-tech signaling method can make some sense, obviating the need for ET to maintain a fully operational radio or laser transmitter over potentially millions or billions of years. The probe might send a pulsed message by successively turning the mirror toward and away from Earth. However, it may be that the mirror is meant only to attract Earth's attention, and that its informational payload may need to be physically retrieved by a space mission.

7. An alien probe entering our Solar System for the purposes of extended surveillance, as opposed to a flyby mission, would need to decelerate into an orbit. A highly eccentric orbit would require the least amount of decelerant energy. A highly eccentric orbit would also allow the probe to surveil the largest amount of the Solar System, its possible preference. If such a probe is intentionally signaling Earth, and since the orbit could be on any inclination, the best place to detect it would be as it passes close to the Sun at perihelion, since that is the only small area of space through which we know some highly eccentric orbits must pass. However, such a probe in a highly eccentric orbit will spend only a very small fraction of its orbital periodicity at perihelion. Consequently, the only way we would likely detect such a probe (unless we persistently observe the Sun) would be if there are very many such alien probes and thereby the chances of any one of them being near perihelion in a given moment proportionally increased. The better reason to observe the area around the Sun is that an alien probe might assume a close Solar orbit. Such a close orbit would not only best allow it to harvest energy, it may allow the probe to be more easily located by a sister probe or node in another star system. The Sun at 1 AU is a strong radio source in itself. However, the signal to noise ratio increases rapidly with distance from it. For example, at 5 solar radii, corresponding to about 0.02 AU, the Green Bank Telescope would lose about 90% of its sensitivity to an ET signal. However, the remaining sensitivity would be adequate to detect anything but an extraordinarily faint signal, and since this is about 20 times closer to the Sun than Mercury's orbit, it is hard to imagine that an ET probe would be that close, except when approaching transit or occultation. The alien nodal network protocol may require that all probes assume tight solar orbits in order to facilitate communication. The ESA Solar Orbiter orbits at 0.5 AU. Presumably, ET can devise materials (e.g., advanced ceramics) that can survive closer orbits yet. Sister probes then need merely transmit a laser beam at the direct center of the star, that beam being just wide enough to illuminate the probe's entire tight orbit. A local probe intentionally signaling Earth would presumably do so with an unmistakably loud or bright signal relative to a putative interstellar beacon. Unless such a signal is either highly intermittent or its inception quite recent, it might already have been observed in the foreground of SETI observations of more remote targets. However, a tight solar orbit would be exempt from this truism since this is a parameter space that has been neglected in all previous SETI searches.

8. In addition to the Moon, as discussed above, an alien probe might be located on any astronomical body within our Solar System. However, geologically active bodies may be less likely bases than geologically inactive bodies. Over deep time, a geologically active body presents risks such as burial in lava. Asteroids would be ideal bodies in that they are often materially rich, close enough to the Sun to utilize its energy, and, except for the largest among them, geologically inactive. Therefore, a program of observing all known asteroids in particular, or the entire ecliptic plane in general, is advised. It is hard to know what materials a probe would most want to exploit, and therefore, whether it would prioritize metals, carbon, or silicon for the purposes of self-repair, build-out capacities, or replication (i.e., create daughter von Neumann probes). That said, the asteroid Psyche should receive particular scrutiny, as this peculiar M-class asteroid is possibly the core of a would-be planet, and contains about one million times more metals than ever mined from the surface of the Earth. The protocol for mosaicking the ecliptic plane is no different in practice from a general mosaicking of the Solar System, except limited to a discrete band above and below the ecliptic plane. The ecliptic can be preferentially targeted where it crosses the plane of the Milky Way to maximize the chances of registering a serendipitous background detection.

9. Benford has recommended targeting objects that are located in several highly-stable near-Earth orbits, namely, co-orbital bodies, quasi-satellites, and a trojan asteroid [15]. Benford argues that these may be better objects to consider than main belt asteroids, because they would give an ET probe a nearby platform from which to surveil Earth, would be closer to the Sun for more efficient energy harvesting, and would be rich in metals and volatiles that a probe may require for repair or build-out. Alternatively, one or more of these objects may themselves be probes and not natural bodies. Earth's only known trojan asteroid, the ~300 m. wide 2010TK7, located at Lagrange point L4, should be especially targeted. 16 other "co-orbital" objects have been identified to date, and many more are presumably awaiting discovery. The potentially greater value suggested by these special areas of space is that ET probes might assume orbits within them that remain stable over millions of years with little or no energy expenditures required on their part. Consequently, the entirety of these co-orbital planes should be observed with wide FOV telescopes, and not just the specific co-orbital objects that have been discovered to date.

10. Earth would seem to be a poor ET probe landing site by the criteria mentioned above, as it is a relatively deep gravity well with a significant atmosphere, requiring that the probe have landing equipment and a heat shield suitable for the task. Also, Earth is geologically and biologically active, posing severe dangers over deep time during which a probe might be subducted, buried in lava, stepped on by a dinosaur, overgrown by a jungle, and so forth. Nonetheless, one must admit that it is the very closest spot from which to surveil Earth. The very possibility of a recent or distant past Earth landing rarely appears in the reputable SETI literature for two main reasons. First, it smacks of highly-charged UFOism. Second, SETI, as a reputable scientific endeavor, currently lies wholly within the domain of astronomy, and not so far taken up by Earth sciences. But there is no law that says that archeologists and geologists cannot get into the hunt. In fact, the Air Force has recently upgraded its investigation of UFOs with the August 4, 2020 announcement of the establishment within the Pentagon of an Unidentified Aerial Phenomena Task Force created by Deputy Defense Secretary David Norquist. Purported sightings by air force pilots of objects that defy all known aerodynamics in their movements leads to the deduction that if the observed phenomena really are objects of alien derivation, they must be AI probes rather than "manned" by biological beings because the latter would presumably be crushed by the G-forces of their very large accelerations.

11. On August 16, 1977 the Big Ear Radio Telescope at Ohio State University recorded a signal that looked very much like a putative interstellar beacon. Jerry Ehman, upon reading the printout of the signal wrote "Wow!" in the margins, and the signal has been known as the "Wow!" signal ever since. The Wow! was never seen again at the same coordinates by either the Big Ear or any other radio telescope (however, although the SETI literature seems to assume that there have been many re-observations, a detailed review indicates that this is not the case [39]). Although the Wow! is by far the most famous example, there have been quite a few other recorded signals that have evinced the presumed qualities of an alien signal save for the fact that they have not been re-observed at the same sidereal coordinates. These signals are known as "transients" in the SETI literature. However, all re-observations to date have worked under the presumption that the purported beacon is located in a distal star system, which would appear stationary on human timescales. However, if the transient derived from a probe within our own solar system it would be in apparent motion. A simple question should be asked of all historical transients: what local object might have been in the FOV at the time of original observation? If, for example, there was an asteroid in the FOV, then that asteroid should be observed in its now current location. Additionally, in the event that the transient was recognized in near real time, the telescope might make successive observations in an outward-moving spiral from the coordinates of the original observation.

## 4 LOCAL METI AS A STRATEGY AND AS A COMPROMISE

There is a subset of SETI scientists and enthusiasts who postulate that it is incumbent upon humans to initiate contact with aliens [18]. In effect, they argue that the galactic-wide communication protocol might require that new civilizations who seek membership in the galactic club make the first request for contact. However, since about 96% of stars are older than Sol, one of the few things we know for a fact about ET is that given that the first civilizations might have arisen at least five billion years ago, it is statistically almost an impossibility that the civilization that we first encounter will, like us, be only in its first century of being able to send or receive radio or optical signals. This leads to the surmise that their capabilities — including the capability to do us harm if they choose — are vastly in advance of our own. Opponents therefore argue that as the very newest technological civilization in the galaxy it is better to listen and learn, rather than speak first and possibly be very sorry. Perhaps the most damning failure of the would-be METI enterprise is a lack of any strategy to receive a return message. The first METI effort was made by Frank Drake, when in 1974 he sent a radio signal from Arecibo to M13, a globular cluster located at a distance of approximately 22,000 light years from Earth. A return message would therefore be expected in about 44,000 years from now. Unfortunately, Drake failed to reserve the exclusive rights to the Arecibo receiver in or about the year 46,000 AD. Consequently, he encumbered future Earthlings with the risk that M13ians might be hostile, while making no provisions to receive their vast wisdom in the event that they are benign. All subsequent METI efforts to date have suffered from the same shortfalling. Even when their targets have been nearby stars, no planning whatsoever has been made to receive a return message in double the time indicated by the target's distance in light years.

However, METI targeting of local Solar System objects may be an acceptable alternative for two reasons:

- METI proponents have argued that it is acceptable to transmit messages to stars because ET must already know, by virtue of having detected our EM leakage, that we are here. However, our omnidirectional radio leakage (often referred to as "I Love Lucy" leakage) would damp down to incoherent noise by the time it reached even nearby stars unless an alien civilization was deploying gargantuan receivers and very long integration times. Therefore, not surprisingly, METIists propose to transmit a directed beam with a power that might be approximately $10^5$ greater than the omnidirectional leakage they claim ET can already decipher. On the other hand, local surveilling probes would presumably be able to detect the leakage and therefore know that we had attained obtained technological competence, so it really would be useless to try to conceal this fact from them.

- Local distances are measured in light minutes or hours, rather than light years. For example, even distant Pluto lies only approximately 280 light minutes (4.33 hours) away. It would be possible to send a message to Pluto early in the evening and receive a message in return by the next dawn. The largest asteroid, Ceres, is currently only about 17 light minutes away.

It is therefore suggested that a reasonable compromise between the METI and anti-METI camps might allow for local METI, while banning interstellar METI. The word, "banning" is used intentionally to mean that the enterprise should be sub-ject to government regulations [40]. For example, local METI would be allowed only below a certain intensity threshold (to diminish the chances that it might illuminate a background star with a decodable flux), would be banned during times when the local target occults nearby stars or the plane of the Milky Way, and would be permitted only when the METIists have demonstrated that they have access to an appropriate receiver at the time that a response might be anticipated. Since the METIists would be speaking on behalf of all humankind, the content of their message should be pre-approved by the UN Security Council and/or the UN General Assembly.

If one accepts the probability that ET is at least as likely to be detected locally as by interstellar beacon, then this seems like a more than reasonable compromise, and not too much to ask of METIists who currently feel empowered to transmit their messages to any target in the universe completely on their own authority.

A compromise means that both parties give up something. Opponents to METI can argue with real justification that even local METI may be dangerous, even when allowing that the alien probe knows that a technological species has emerged on Earth. A probe's reception of a directed transmission from Earth will indicate to it that it has been detected, an event that might "frighten it" (albeit, we are referring to an AI entity), causing it to lash out defensively. Consequently, even local METI should be undertaken with the consent of humankind as expressed through the U.N.

## 5 CONCLUSIONS

For the better part of 60 years, SETI scientists have assumed that an alien signal will derive from far distant interstellar coordinates. Due to the inverse square law, that signal would likely be very weak and would therefore be best detected with a highly sensitive telescope, which in turn would likely have a small FOV. A small FOV is desirable when the purported source is a star. Even then, in the radio spectrum, it is not expected that even our most sensitive telescopes will be able to detect more than an alien carrier wave. With either a large optical telescope or large radio dish, the only way that a large number of stars can be sampled at once is to point toward another galaxy, or to look deep within the plane of the Milky Way, stacking stars depth-wise.

Testing the alien probe hypothesis requires a significant shift in strategy. Because a local signal will come from a distance that is many orders of magnitude closer to Earth than an interstellar signal, even a modest local transmitter will vastly outshine an interstellar signal, unless the latter transmits at an almost unimaginable intrinsic flux. Consequently, the large aperture and small FOV of large telescopes are counter-productive to a local search.

It may be that there are multiple local nodes, each located on lines extending outward from 550 AU [26, 27] and each pointed towards different nearby stars, forming part of a very complex galactic communication system. Because each leg of the communications network would be short, narrow beamed lasers would best suit the purpose. Because those beams are narrow and only connect the local nodes with nearby stars, they are under most circumstances invisible to Earth, giving the impression that there is a "Great Silence," when in fact Earth may be embedded within a rich cacophony.

## Acknowledgements

I would like to thank Geoff Marcy and Andrew Siemion for their help in reviewing this paper.